\documentclass[twocolumn, prb, showpacs]{revtex4}
\usepackage{graphicx}
\usepackage{dcolumn}
\usepackage{bm}
\usepackage{color}

\begin{document}


\title{Two-step antiferromagnetic transition and moderate triangular frustration in Li$_2$Co(WO$_4$)$_2$}

\author{I. Panneer Muthuselvam$^{1}$, R. Sankar$^{1}$, A. V. Ushakov$^{2}$,  G. Narsinga Rao$^{1}$, Sergey V. Streltsov$^{2,3}$ and  F. C. Chou$^{1,4,5}$}
\email[corresponding author: ]{fcchou@ntu.edu.tw}
\affiliation{$^{1}$Center for Condensed Matter Sciences, National Taiwan University, Taipei 10617, Taiwan}
\affiliation{$^2$Institute of Metal Physics, Russian Academy of Science, S. Kovalevskaya Str. 18, 620041 Ekaterinburg, Russia}
\affiliation{$^3$Ural Federal University, Mira St. 19, 620002 Ekaterinburg, Russia}
\affiliation{$^4$National Synchrotron Radiation Research Center, Hsinchu 30076, Taiwan}
\affiliation{$^5$Taiwan Consortium of Emergent Crystalline Materials, Ministry of Science and Technology, Taipei 10622, Taiwan}

\date{\today}

\begin{abstract}
We present a detailed investigation of the magnetic properties of the spin-$\frac{3}{2}$ system Li$_2$Co(WO$_4$)$_2$ by means of magnetic susceptibility and specific heat. Our experimental results show that in Li$_2$Co(WO$_4$)$_2$
a short-range antiferromagnetic (AFM) correlations appear near $\chi$$_{max}$ $\sim$ 11 K and two successive long-range AFM phase transitions are observed at T$_{N1}$$\sim$ 9 K and T$_{N2}$$\sim$ 7 K. The frustration factor, $\mid$$\Theta$$\mid$/T$_{N1}$$\sim$3, indicates that the system is moderately frustrated, which is identifiable by the broken triangular symmetry within both $ab$- and $bc$-planes for the triclinic crystal structure. The magnetic isotherm at temperatures below T$_{N2}$ shows a field-induced spin-flop transition, and a complete H-T phase diagram for the two-step AFM system is mapped.  $Ab$~$initio$ band structure calculations suggest that the strongest exchange coupling does not correspond to the shortest Co-Co distance along the $a$-axis, but rather
along the diagonal direction through a Co-O-W-O-Co super-superexchange
path within the $bc$-plane.


\end{abstract}


\maketitle

\section{\label{sec:level1} Introduction\protect\\ }

Although one-dimensional (1D) antiferromagnetic (AFM) systems are not expected to show spin long-range ordering (LRO) as a result of strong quantum fluctuations, 3D AFM LRO has been observed in most quasi-1D spin chain
compounds because of the weak, but non-zero interchain couplings.  For gapped quasi-1D compounds, such as PbNi$_2$V$_2$O$_8$, SrNi$_2$V$_2$O$_8$, TlCuCl$_3$ and Ni(C$_5$H$_{14}$N$_2$)$_2$N$_3$(PF$_6$), a strong magnetic field can destroy the gap and lead to the formation of an AFM ground state at low temperatures.~\cite{Tsujii2005,Bera2013,Oosawa1999,Honda1998}
Cobalt-based, low-dimensional magnetic systems often demonstrate
behaviors of spin-flop and field-induced order-disorder
transitions.~\cite{He2005,He2006,He2009,Mohapatra2009,Markina2014,Yokota2014}
The most fascinating characteristic of low dimensional systems is the observation of magnetization plateaus at high field, i.e., magnetization could stabilize at a fraction of the saturated magnetization in quantum nature.\cite{Honecker2004}
Two-step successive magnetic phase transition has been observed in 1D and 2D Co spin systems, such as  Pb$_3$TeCo$_3$V$_2$O$_{14}$ and Ba$_3$CoNb$_2$O$_9$.~\cite{Markina2014,Yokota2014}  The common feature found in these two compounds is the persistent triangular symmetry of the Co spin network: either triangular tubing or planes.  The comparative study between Ba$_3$CoNb$_2$O$_9$ and Ba$_3$CoSb$_2$O$_9$ suggests that the 2D triangular-lattice antiferromagnets (TLAF) of uniaxial anisotropy exhibits a two-step magnetic phase transition, whereas a single transition takes place in systems with easy-plane anisotropy.~\cite{Yokota2014, Susuki2013}

\begin{figure}
\begin{center}
\includegraphics[width=3.5in]{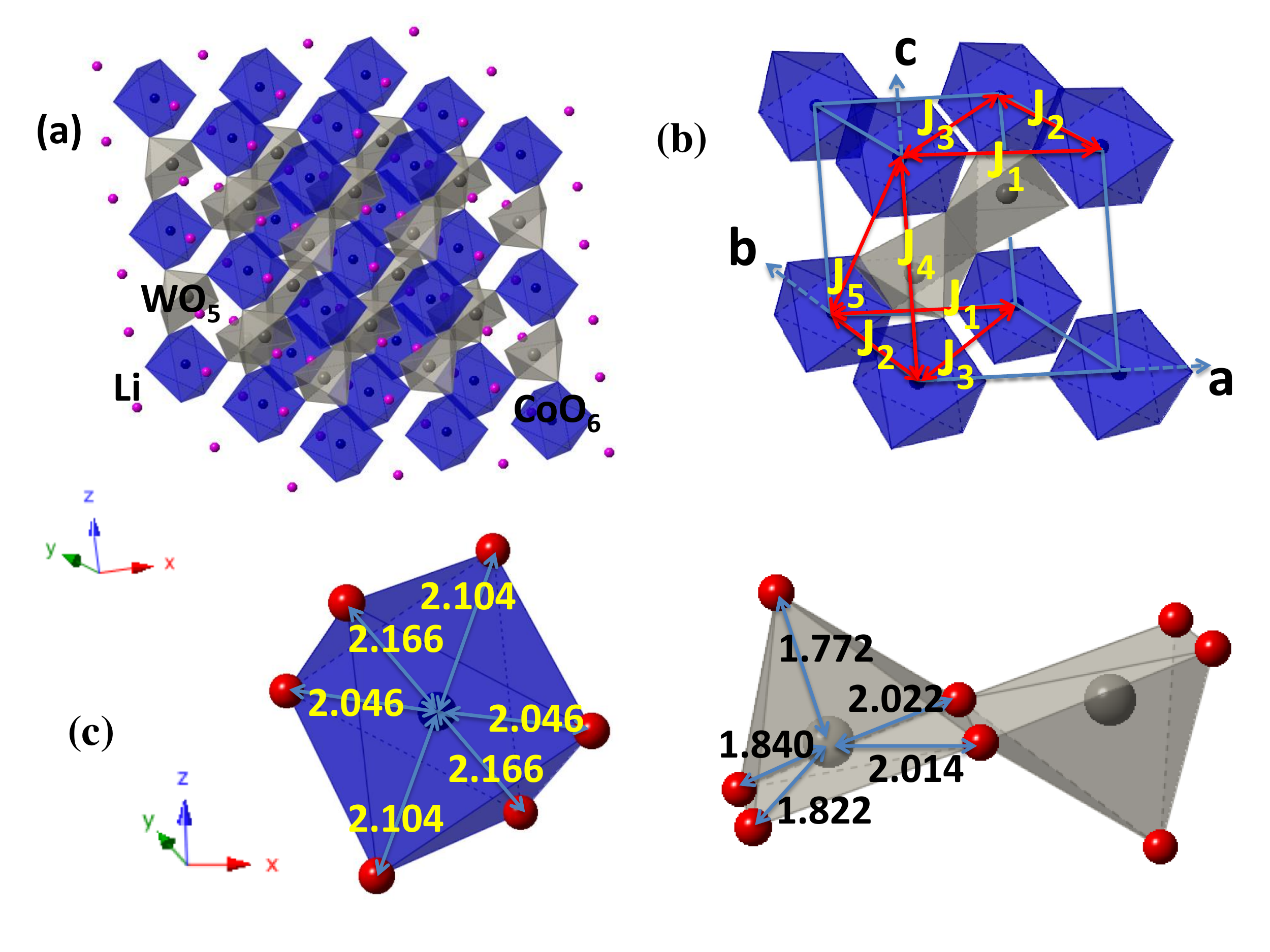}
\end{center}
\caption{\label{fig-structure}(color online)(a) Triclinic crystal structure of Li$_2$Co(WO$_4$)$_2$ shown with the CoO$_6$ octahedra in blue and the 
WO$_5$ pyramidal pairs in grey. (b) Magnetic exchange interactions (J$_1$-J$_5$) 
between the Co spins are shown, and the corresponding Co-Co distances are summarized in Table I.  (c) Bond lengths in $\AA$ for the CoO$_6$ octahedra and the WO$_5$ 
pyramids.}
\end{figure}

Li$_2$Co(WO$_4$)$_2$ crystalizes in a triclinic crystal structure of space group $P\bar{1}$, as shown in Fig.~\ref{fig-structure}. The CoO$_6$ octahedra are corner-shared with pairs of the WO$_5$ edge-shared pyramids, and Li ions are located in the interstitial sites. The Co-Co distances and Co-W-Co angles are summarized in Table~\ref{Table2} for the unit cell of Li$_2$Co(WO$_4$)$_2$ shown in Fig.~\ref{fig-structure}(b).
The CoO$_6$ octahedron is slightly distorted as indicated by the Co-O distances of 2.047 $\AA$, 2.099 $\AA$, and 2.164 $\AA$.
The WO$_5$ pyramids are inverted within each edge-sharing pair as shown in Fig.~\ref{fig-structure}(c).



Recently, we found that Li$_2$Co(WO$_4$)$_2$ also possesses a two-step successive AFM transition similar to that of the examples cited above.
To better understand the origin of 
this two-step successive magnetic transition, correlations to crystal dimensionality, spin anisotropy, triangular symmetry breaking, and the role of Dzyaloshinskii-Moriya interactions should all be carefully examined.   The triclinic crystal structure of Li$_2$Co(WO$_4$)$_2$ provides a rare opportunity to investigate the origin of successive phase transitions found in the low dimensional cobaltate system, where Co spins could be viewed as two quasi-triangles along two crystal axes.

In this report, we present the investigation of the magnetic properties of Li$_2$Co(WO$_4$)$_2$.  
Short-range AFM correlations were found to 
result in the formation of the broad peak of $\chi_{max}$(T) at T$\sim$ 11 K, followed by a two-step
AFM-like magnetic transition at T$_{N1}$$\sim$ 9 K and T$_{N2}$$\sim$ 7 K. A two-step field-induced spin-flop transition was also observed in the magnetization isotherms below T$_{N2}$.
Finally, we constructed a schematic phase diagram based on the results of the magnetic and specific
heat measurements and found main exchange interaction parameters using band structure calculations.

\section{Experimental details}

Polycrystalline Li$_2$Co(WO$_4$)$_2$ powder was prepared by a conventional solid state reaction. Stoichiometric amounts of high purity ($>$99.95\%) CoO, Li$_2$CO$_3$, and WO$_3$ were mixed and ground homogeneously using a mortar and pestle. The homogenized mixture of oxides was heated at 550$^\circ$C for 24 hours in air. The calcined powder sample was pressed into pellets and heated up to 600$^\circ$C for 24 hours and then 650$^\circ$C for 160 hours in air with several intermediate grindings. The phase purity and structure refinement were confirmed by powder diffraction using synchrotron X-rays of $\lambda = 0.619 $ \AA$ $ (NSRRC, Taiwan) at room temperature.  The Rietveld refinement of the SXRD pattern of the sample could be indexed to a triclinic crystal structure with the space group P$\bar{1}$ without any impurity phase. The refined lattice parameters of Li$_2$Co(WO$_4$)$_2$ are a = 4.90724(7) $\AA$, b = 5.61876(8) $\AA$, c = 5.86495(8) $\AA$, $\alpha$=70.720(1)$^\circ$, $\beta$=88.542(1)$^\circ$, and $\gamma$=115.479(1)$^\circ$, and V = 135.110(3)$\AA^3$, which are in good  agreement with those reported in the literature.\cite{Vega2001}
These structural parameters were used in the {\it ab initio} band structure calculations.
 The dc magnetization was measured using a SQUID VSM (Quantum Design, USA) under zero-field-cooled (ZFC) and field-cooled (FC) conditions. The heat capacity was measured using a standard relaxation method with a PPMS (Quantum Design, USA).

\section{Calculation details}

The crystal structure data for the $ab$-$initio$ band structure calculations
were taken from the above mentioned refinement results. The linearized muffin-tin orbitals method (LMTO)~\cite{Andersen1984} was used in the calculations with the von Barth-Hedin exchange correlation potential.~\cite{Barth-1978} The strong
Coulomb interaction in the $3d$-shell of the Co$^{2+}$ ions was taken into account within the
LSDA+$U$ method.~\cite{Anisimov1997} The on-site Coulomb repulsion, $U$, and the intra-atomic
Hund's exchange parameter, $J_H$, were chosen to be $U$(Co)$=7$ eV and $J_H$(Co)$=0.9$
eV.~\cite{Nomerovannaya2004,Markina2014}
We used a mesh of $96$ $k$-points in the full Brillouin zone during the
course of the calculations.

The exchange coupling integrals, $J$, were calculated for the Heisenberg
model written as
\begin{equation}
\label{heis}
H = \sum_{ij} J_{ij} \vec{S_i}\vec{S_j},
\end{equation}
where the summation runs twice over each pair. We utilized the Liechtenstein's exchange interaction
parameters (LEIP) calculation procedure, where $J$ is determined as a second derivative of the energy with
respect to a small spin rotation.~\cite{Katsnelson2000}
The spin-orbit coupling was not taken into account. 

The Co$^{2+}$ ions form 
triangles along the \textit{ab} and \textit{bc} planes, as shown in Fig. \ref{fig-structure} (b). 
The Co-Co bond lengths at room temperature for the Co quasi-equilateral triangular unit within the \textit {ab}-plane are 4.902~$\AA$,
5.650~$\AA$ and 5.618~$\AA$. Similarly, the Co triangular unit within the \textit {bc} plane has slightly longer Co-Co distances of 5.618~$\AA$, 5.865~$\AA$ and 6.648~$\AA$, as seen in Fig.~\ref{fig-structure}(b). The other Co-Co distances are much larger, thus the Heisenberg exchange interaction integrals were only estimated for the aforementioned bonds.
For the calculations, we used the AFM structure in which the magnetic moments for the 4.902~$\AA$ and
5.618~$\AA$ Co-Co bonds were ordered antiferromagnetically,
whereas those for the 5.65~$\AA$ Co-Co bond  
ferromagnetically. We also checked that the use of other AFM configurations does not change the calculation results (i.e., the $J$ values). Spins along the
$c$-axis were taken to be antiferromagnetically ordered.

\section{\label{sec:level1} RESULTS AND DISCUSSION \protect\\ }


\begin{figure}
\begin{center}
\includegraphics[width=3.5in]{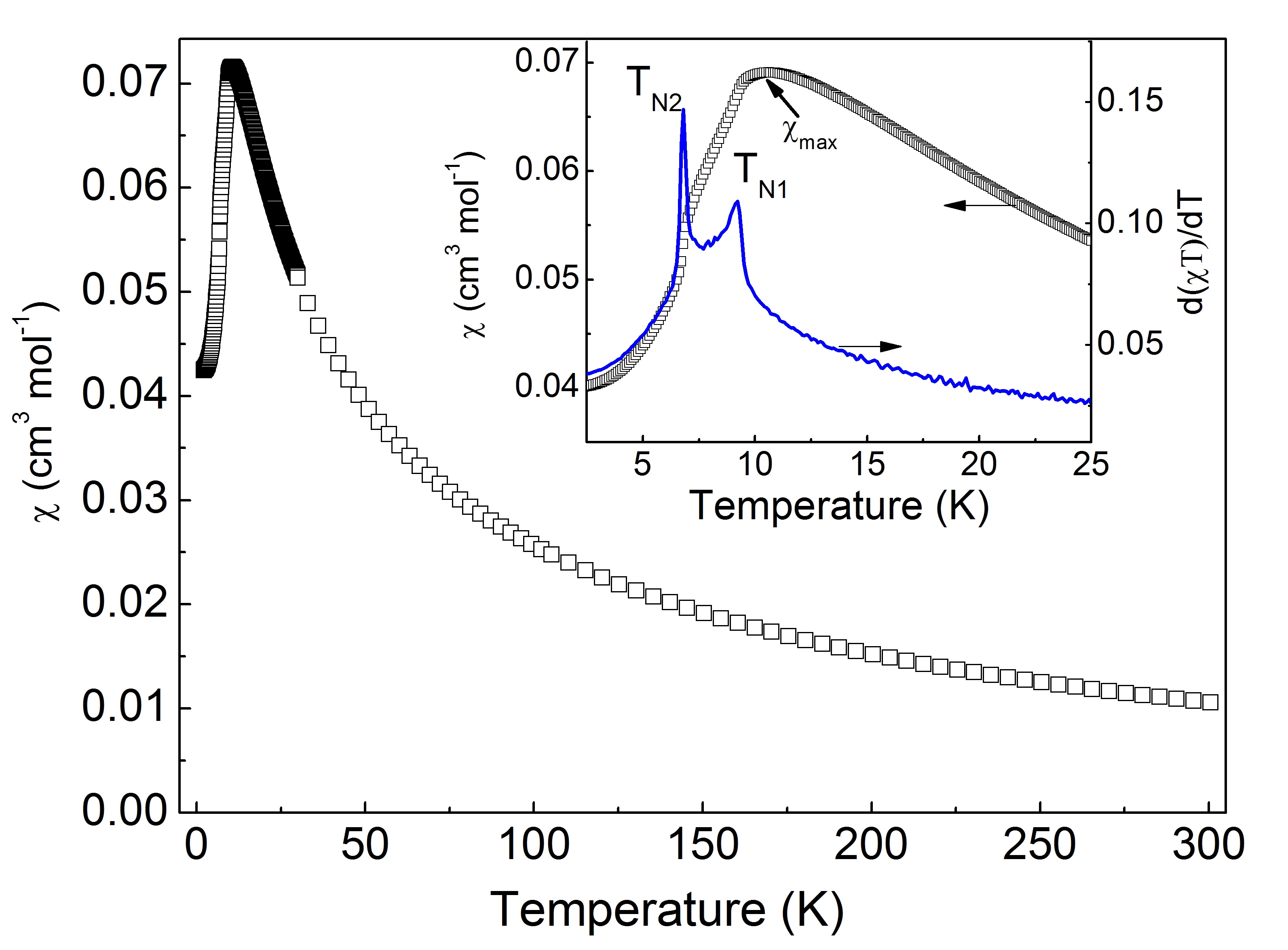}
\end{center}
\caption{\label{fig-chiT}(color online) Temperature dependence of the magnetic susceptibility measured under an applied magnetic field of 1 T. The left axis of the inset highlights the low temperature regime and the right axis highlights the derivative, d($\chi$T)/dT.}
\end{figure}

\begin{figure}
\begin{center}
\includegraphics[width=3.5in]{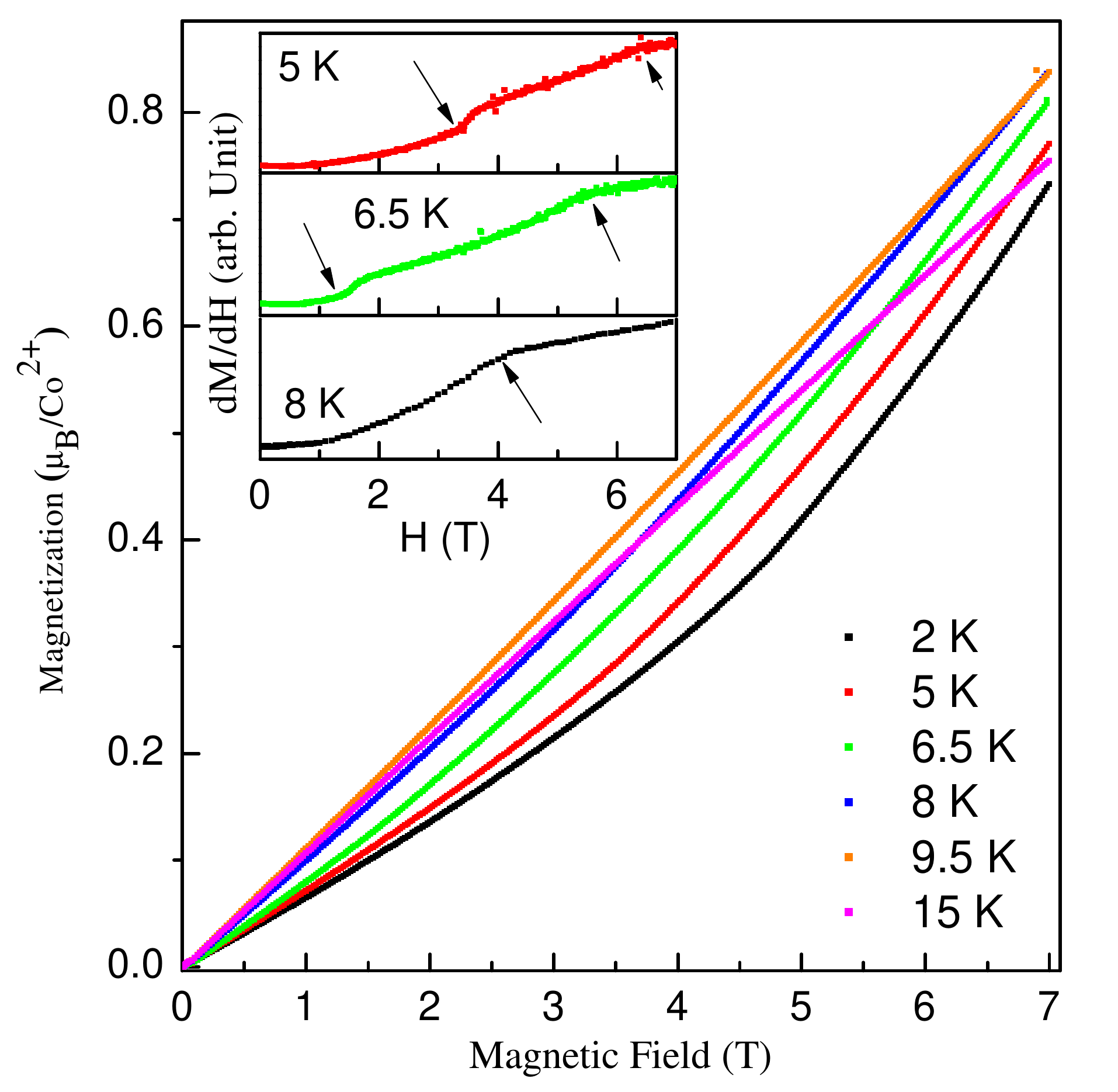}
\end{center}
\caption{\label{fig-MH}(color online) Field dependence of the magnetization measured at different temperatures for Li$_2$Co(WO$_4$)$_2$. The inset shows the derivative the of magnetization with respect to magnetic field. }
\end{figure}

\subsection{\label{sec:level1} Magnetic Susceptibility \protect\\ }

Fig. \ref{fig-chiT} shows homogeneous spin susceptibilities $\chi$(T) as a function of temperature measured under an applied magnetic field of 1 T for Li$_2$Co(WO$_4$)$_2$.  No hysteresis was observed between the ZFC and FC measurement data.  The $\chi$(T) shows Curie-like behavior at high temperature and reaches a rounded maximum at approximately $\chi$$_{max}$ $\sim$ 11 K, as shown in the inset of Fig. \ref{fig-chiT}, which indicates the characteristic behavior of short-range AFM correlation for a low dimensional spin system.  Moreover, small but sharp drops in the $\chi$(T) at approximately 9 and 7 K can be identified more clearly by the d($\chi$T)/dT shown in the inset of Fig. \ref{fig-chiT}, at T$_{N1}$ $\sim$ 9 K and T$_{N2}$$\sim$ 7 K.  The preliminary neutron powder diffraction measurement results indicate that these phase transitions 
can be attributed to paramagnetic(PM)-to-incommensurate (IC) AFM long-range ordering (LRO) at T$_{N2}$ $\sim$ 9 K and incommensurate(IC)-to-commensurate (CM) AFM LRO at T$_{N1}$ $\sim$ 7 K.\cite{Sunil2014}


The $\chi$(T) follows Curie-Weiss behavior well above 130 K, and the Curie-Weiss law fitting of $\chi=\chi_\circ + C/(T-\Theta)$
 indicates a Curie constant of C= 3.48 cm$^3$ K/mol and Curie-Weiss constant of $\Theta$ = -27 K. The negative value of $\Theta$ suggests that the effective exchange interactions between Co$^{2+}$ ions are AFM. The effective magnetic moment ($\mu_{eff}$) was calculated to be $\sim$ 5.27 $\mu_B$, which is much larger than the expected S=3/2 spin-only value of $\sim$ 3.87 $\mu_B$ for Co$^{2+}$ ($3$d$^7$ configuration in high spin state). The high value of the effective magnetic moment suggests the unquenched spin-orbit coupling for the high spin state of Co$^{2+}$.  The obtained $\mu_{eff}$ 
$\sim$ 5.2 $\mu_B$ 
is  similar to those reported for several other Co$^{2+}$-based compounds.\cite{Viola2003,Nakayama2013,He2005CM} The obtained g value derived from the Curie-Weiss constant is 
 2.72.

 The spin frustration ratio, $f$ = $\mid\Theta\mid$/T$_{N1}$, was found to be $\sim$ 3, which is indicative of moderate frustration taking place in the present system.  Moderate frustration ($f$~$\sim$~$2.8$) has been observed in Pb$_3$TeCo$_3$V$_2$O$_{14}$, which 
is characterized by a similar spin lattice topology 
in addition to the resembling two-step successive AFM transition.\cite{Markina2014} Based on the quasi-equilateral triangular symmetry of the Co sublattices within both $ab$- and $bc$-planes (see Fig.~\ref{fig-structure}(b)), it is reasonable to 
expect a moderate frustration among the Co spins if Ising-like anisotropy is considered on either plane.

Fig. \ref{fig-MH} shows the magnetization as a function of applied magnetic field measured at different temperatures. The M(H) curves show nonlinear behavior with the applied magnetic field. At temperatures below T$_{N2}$, an abrupt change of slope in the magnetization is observed above certain critical magnetic fields (shown in inset of Fig. \ref{fig-MH}). Interestingly, a two-step behavior is observed in the magnetization isotherms below T$_{N2}$, which is strongly evidenced by the dM/dH derivative shown in the inset. The change of dM/dH slope 
could be attributed to the field-induced magnetic phase transition, or the spin-flop reorientation.  Based on the slope increase of M(H) above the critical fields, the first transition 
can be attributed to a spin-flop transition in which the magnetic field overcomes 
 the spin anisotropy and the AFM spins flop.  Judging from the M(H) slope decrease in the second step for M(H) below T$_{N2}$ and the first step for M(H) above T$_{N2}$, the field-induced transition cannot be assigned to a spin-flop transition. No hysteresis or remnant magnetization is observed in zero field.  Because it is difficult to determine the easy axis in the AFM state 
using polycrystalline samples, and the magnetization does not saturate up to 7 T, high field experiments on a single crystal sample are necessary for the detailed analysis, especially for the second step observed in M(H).

\begin{figure}
\begin{center}
\includegraphics[width=3.5in]{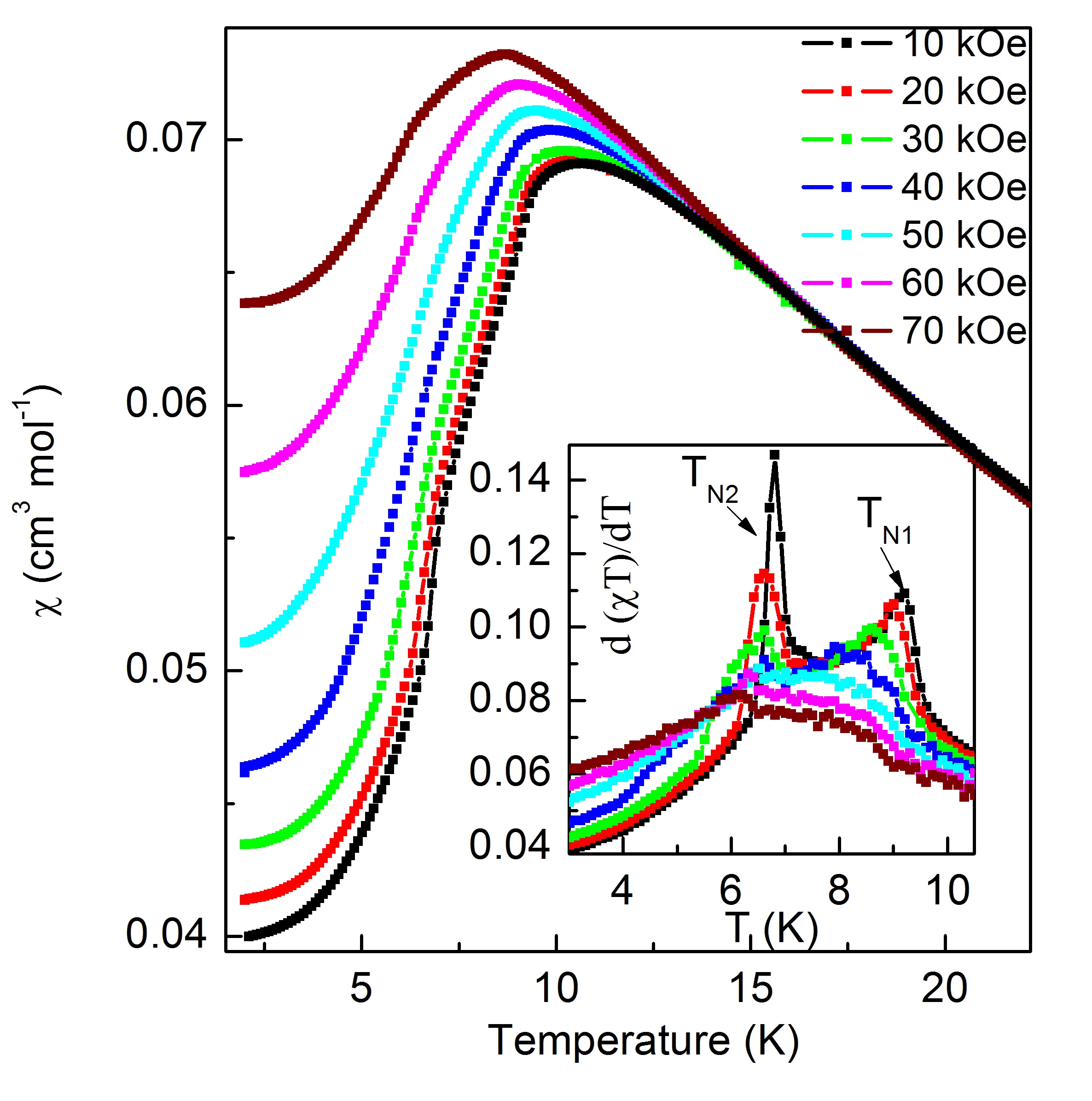}
\end{center}
\caption{\label{fig-chif}(color online) Temperature dependence of the magnetic susceptibility at various applied magnetic fields with d($\chi$T)/dT shown in the inset. }
\end{figure}

To  further clarify the field-induced transition, we have performed magnetic susceptibility measurements in different applied magnetic fields (H). Fig.~\ref{fig-chif} shows the temperature dependence of the magnetization under different fields, and derivative d($\chi$T)/dT is presented in the inset. With increasing field, $\chi$$_{max}$ shifts to lower temperatures. In addition, both T$_{N2}$ and  T$_{N1}$ shift to lower temperatures (inset of Fig. \ref{fig-chif}) at higher fields, although the peaks in d($\chi$T)/dT become significantly broader at high fields.  The decrease of T$_{N2}$ and  T$_{N1}$ is expected for a 3D TLAF with easy-axis anisotropy, when both interlayer and intralayer AFM exchange interactions are of the same order of magnitude.\cite{Plumer1988,Plumer1989} Recently, similar behavior has also been observed in Ba$_3$CoNb$_2$O$_9$.\cite{Yokota2014} However, both T$_{N1}$ and T$_{N2}$ increase with increasing magnetic field when H $\perp$ c in CsNiCl$_3$, where the AFM interlayer exchange interaction is much larger than the intralayer magnetic coupling.\cite{Johnson1979}  It is clear that the magnetic field is effective for lifting the degeneracy of the spin structure with a frustrated triangular geometry.  The Li$_2$Co(WO$_4$)$_2$ with triclinic crystal symmetry possesses a Co spin lattice with a quasi-equilateral triangular geometry, which could explain the reason why the onset of the AFM is so sensitive to the applied field.


\subsection{\label{sec:level1} Specific Heat \protect\\ }

\begin{figure}
\begin{center}
\includegraphics[width=3.5in]{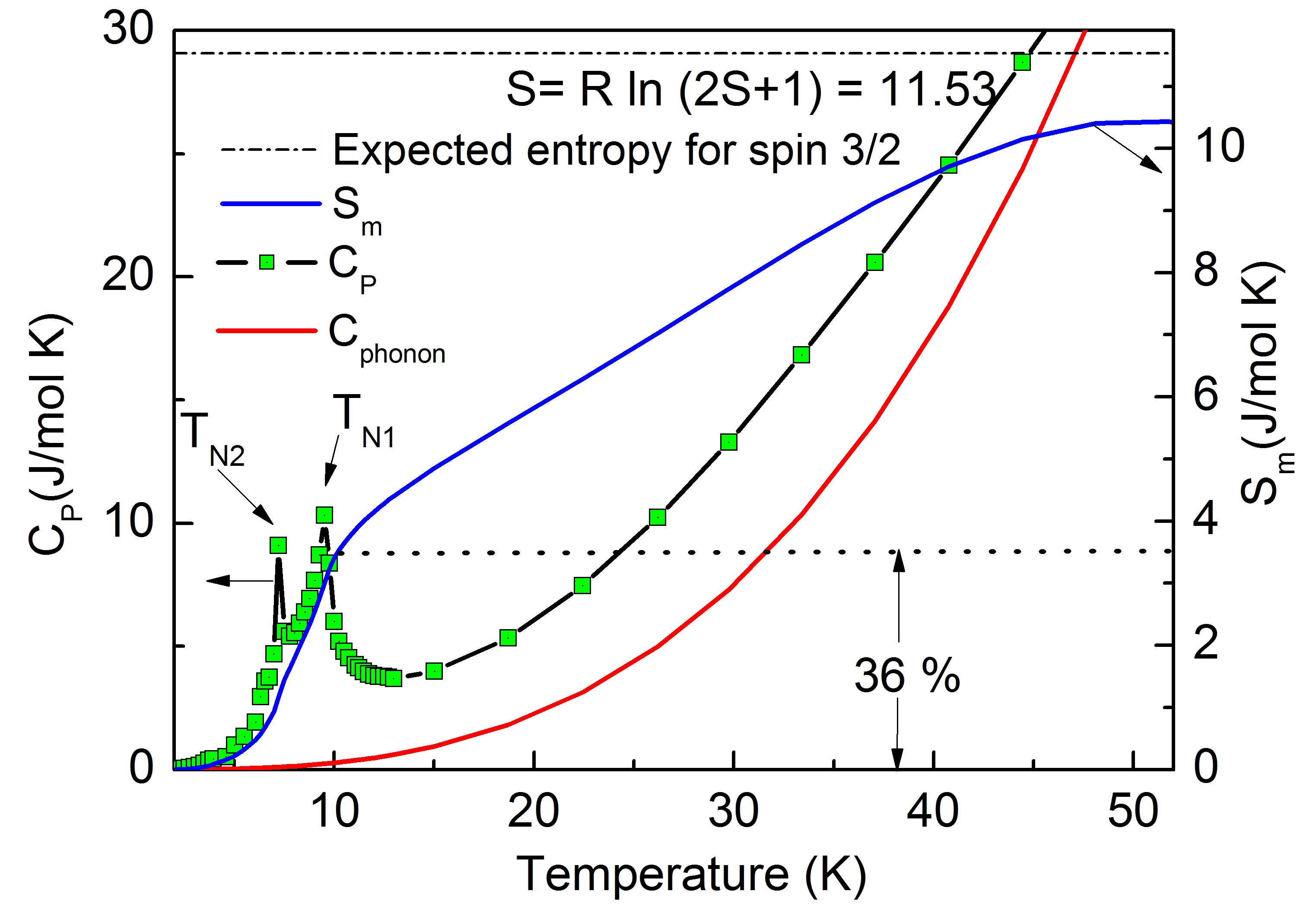}
\end{center}
\caption{\label{fig-cp}(color online) Specific heat as a function of temperature (left, y-axis), and magnetic entropy, S$_m$, versus T (right, y-axis). }
\end{figure}

Fig. \ref{fig-cp} shows the specific heat, C$_P$, as a function of temperature in zero field, where two anomalies are observed at T$_{N1}$ $\sim$ 7.2 K and T$_{N2}$ $\sim$ 9.5 K. These results provide strong evidence that the two successive phase transitions are both of 3D LRO and are consistent with the  AFM nature  inferred from the spin susceptibility measurements. The C$_P$(T) data are fitted using C$_P$/T =$\alpha$ + $\beta$T$^2$ above 15 K to estimate the lattice and spin contributions, which yields $\alpha$ = 0.127 J mol$^{-1}$K$^{-2}$ and $\beta$=2.84x10$^{-4}$J mol$^{-1}$K$^{-4}$.
Because Li$_2$Co(WO$_4$)$_2$ is an insulator, $\alpha$ is assumed to originate from magnetic contributions only. The magnetic specific heat (C$_m$) is obtained after subtracting the lattice contribution ($\beta$T$^3$) from the total C$_P$. C$_m$ is found to follow a linear dependence of T$^3$ below T$_{N2}$, which is indicative of AFM magnon excitations in the ordered state.\cite{Gotaas1985} The magnetic entropy (S$_m$) is estimated through integrating C$_m$/T as a function of temperature, as shown in Fig. \ref{fig-cp}. The S$_m$ increases with increasing temperature and saturates to 10.45 J/mol K above $\sim$52 K, which suggests the measured saturated entropy is approximately  91$\%$ of the calculated value of the total spin entropy of R$ln$(2S+1)= 11.53 J/mol K (R = 8.314 J/K mol) for spin =3/2. However, the entropy gain at T$_{N1}$ is only $\sim$1/3 of the total spin entropy for spin 3/2, which suggests that Li$_2$Co(WO$_4$)$_2$ could be characterized as a quasi low dimensional spin system, i.e., the remaining spin entropy has to be gradually acquired by the short-range magnetic correlation in a wide range of temperatures above T$_{N1}$.

\begin{figure}
\begin{center}
\includegraphics[width=3.5in]{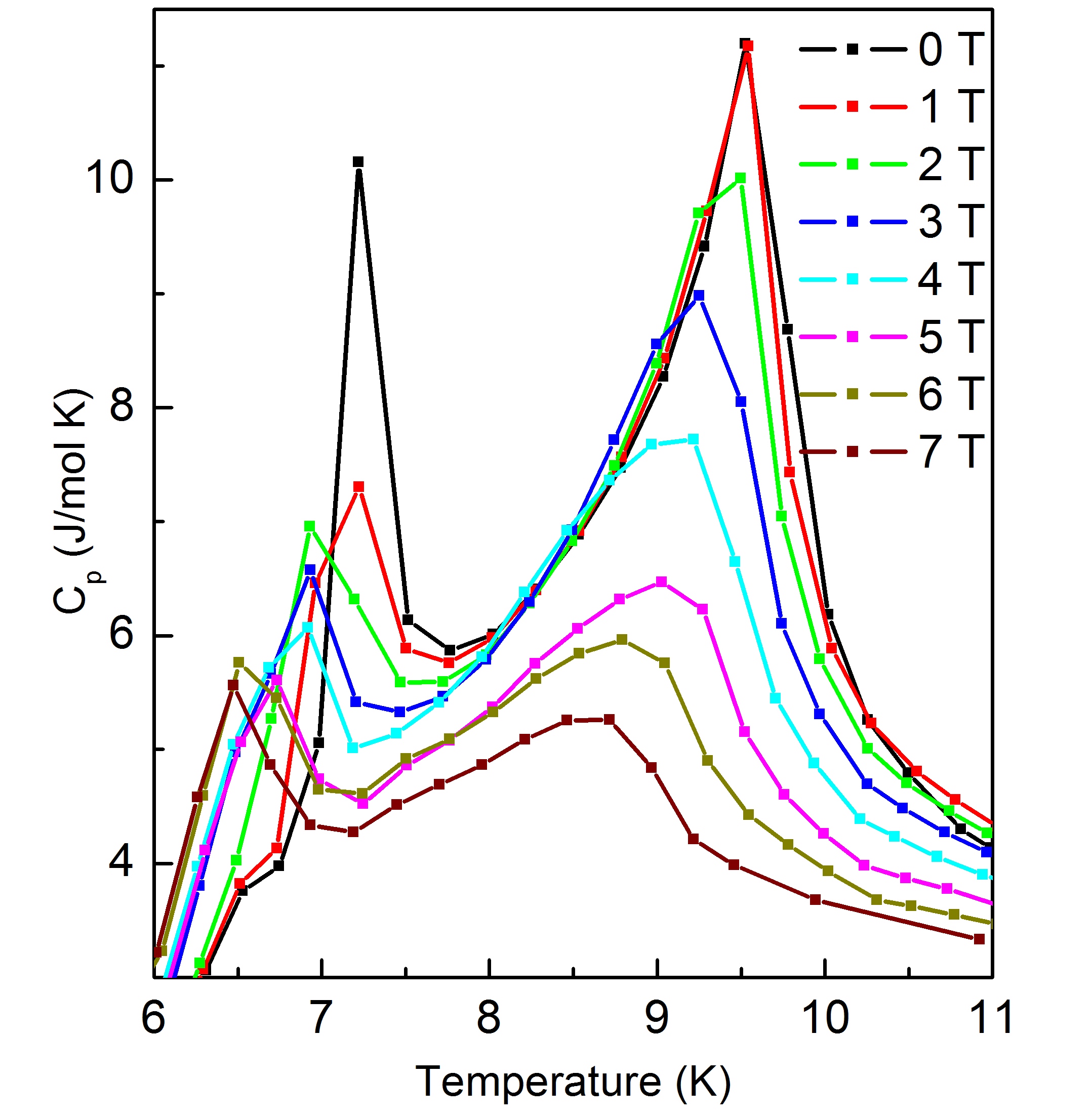}
\end{center}
\caption{\label{fig-cpf}(color online) C$_P$ versus T for different external magnetic fields. }
\end{figure}

C$_P$ measurements in different applied fields have been performed and are shown in Fig. \ref{fig-cpf}. With increasing H, both T$_{N1}$ and T$_{N2}$ peaks shift toward lower temperatures. In addition, the amplitude of the peaks also decreases with increasing H. These results are consistent with the common features of AFM magnetic systems. Although the T$_{N1}$ peak becomes dramatically rounded with increasing H above $\sim$4 T, the T$_{N2}$ peaks are not completely suppressed up to 70 kOe, and no significant broadening occurs except for the 1/2 to 1/3 intensity reduction. The broadening of T$_{N1}$ under high fields suggests that the magnetic field could suppress the inter-plane coupling for the original 3D LRO through moderate frustration, which has also been reflected in the magnetic susceptibilities shown in Fig.~\ref{fig-chif}.  From this point of view, the T$_{N1}$ peak broadening in the field (above T$_{N2}$) could be attributed to a moderate field-induced order-disorder transition. This result was also reflected in the $\sim$1/3 spin entropy gain at T$_{N1}$ (see Fig. \ref{fig-cp}), similar to that obtained in the quasi-1D system of BaCo$_2$V$_2$O$_8$.\cite{He2005}  However, the persistently sharp character of T$_{N2}$ in high fields suggests that LRO 
exists below T$_{N2}$, although it is suppressed at a lower onset.

\subsection{\label{sec:level1} $Ab$~$initio$ calculations \protect\\ }

\begin{figure}
\begin{center}
\includegraphics[width=0.99\linewidth]{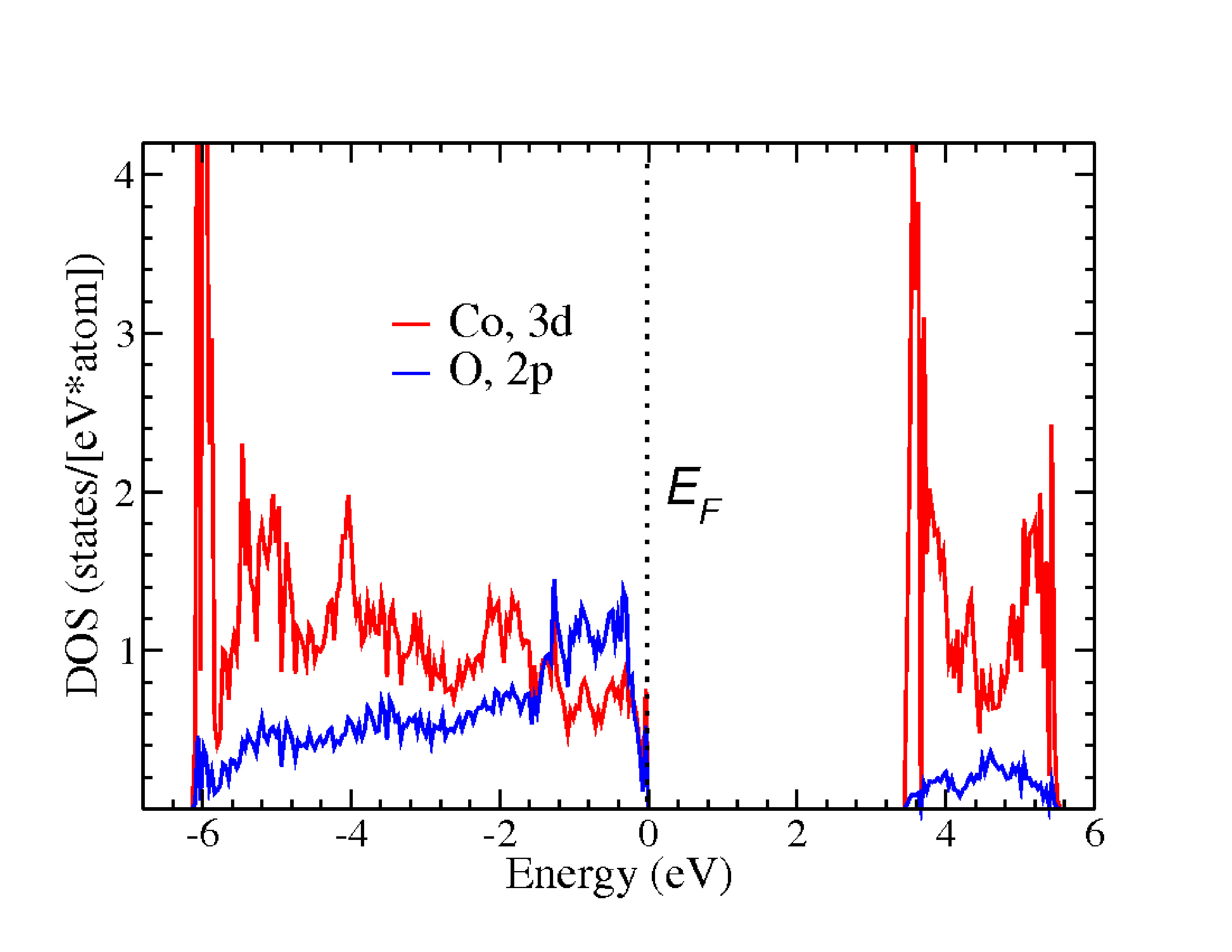}
\end{center}
\caption{\label{tot-dos}(color online) The total densities of Co-$d$ and O-$p$ states in Li$_2$Co(WO$_4$)$_2$. The Fermi energy is in zero.}
\end{figure}

In the LDA+$U$ calculations, Li$_2$Co(WO$_4$)$_2$ was found to be an
insulator with an energy gap of $E_g \sim 3.5$~eV. The electronic density of
states (DOS) are shown in Fig.~\ref{tot-dos}. The top of the valence band is formed by the Co-$3d$ and O-$2p$ states, whereas
the bottom of the conduction band mostly has Co-$3d$ character.
The width of the valence band is quite large, $\sim$~$6$~eV.
The local magnetic moments on the Co$^{2+}$ were found to be $2.75$~$\mu_B$, which are in agreement with 3d$^7$ high spin configuration.

The calculated exchange coupling parameters for Li$_2$Co(WO$_4$)$_2$ are presented in
Table~\ref{Table2}. We estimated the Curie-Weiss temperature as
$\Theta = \frac{2}{3}\sum_i J_i S(S+1)$, where
$S=\frac{3}{2}$ for the spin moment of the Co$^{2+}$ ion, and found that $\Theta =-24.8$~K.
This value is in good agreement with the experimental estimation of $\Theta =-27$~K
given above, which shows that the results of the calculations are reliable.


Based on the coupling constants estimated above, the largest constant
($J_5$) is approximately twice as larger as $J_3$. Hence the system could be considered a set of coupled spin chains, running along the $J_5$ direction as shown in Fig.~\ref{fig-structure}(b) and Table~\ref{Table2}.
Two substantial exchange integrals, $J_3$ and $J_5$, correspond to Co-Co
distances of $5.650$~$\AA$ and $6.648$~$\AA$, respectively.
It is interesting to find that the strongest exchange coupling, $J_5$, does not
correspond to the shortest Co-Co bond lengths, but rather the longest among $J_i$, and the largest bond angle between Co-W-Co.

Each CoO$_6$ octahedron in Li$_2$Co(WO$_4$)$_2$ at room temperature is slightly
elongated along one of the directions, as shown in Fig.~\ref{fig-structure}(c).
This removes the degeneracy in the $t_{2g}$ shell and
 one of the $t_{2g}$ orbitals ($xy$ in the local coordinate system, where $z$ corresponds to the longest Co-O bond) goes higher in energy.
As a result, the $xy$, $3z^2-r^2$ and $x^2-y^2$ orbitals are half-filled and magnetically active for the Co$^{2+}$ ions in the HS configuration.

\begin{figure}
\begin{center}
\includegraphics[width=0.99\linewidth]{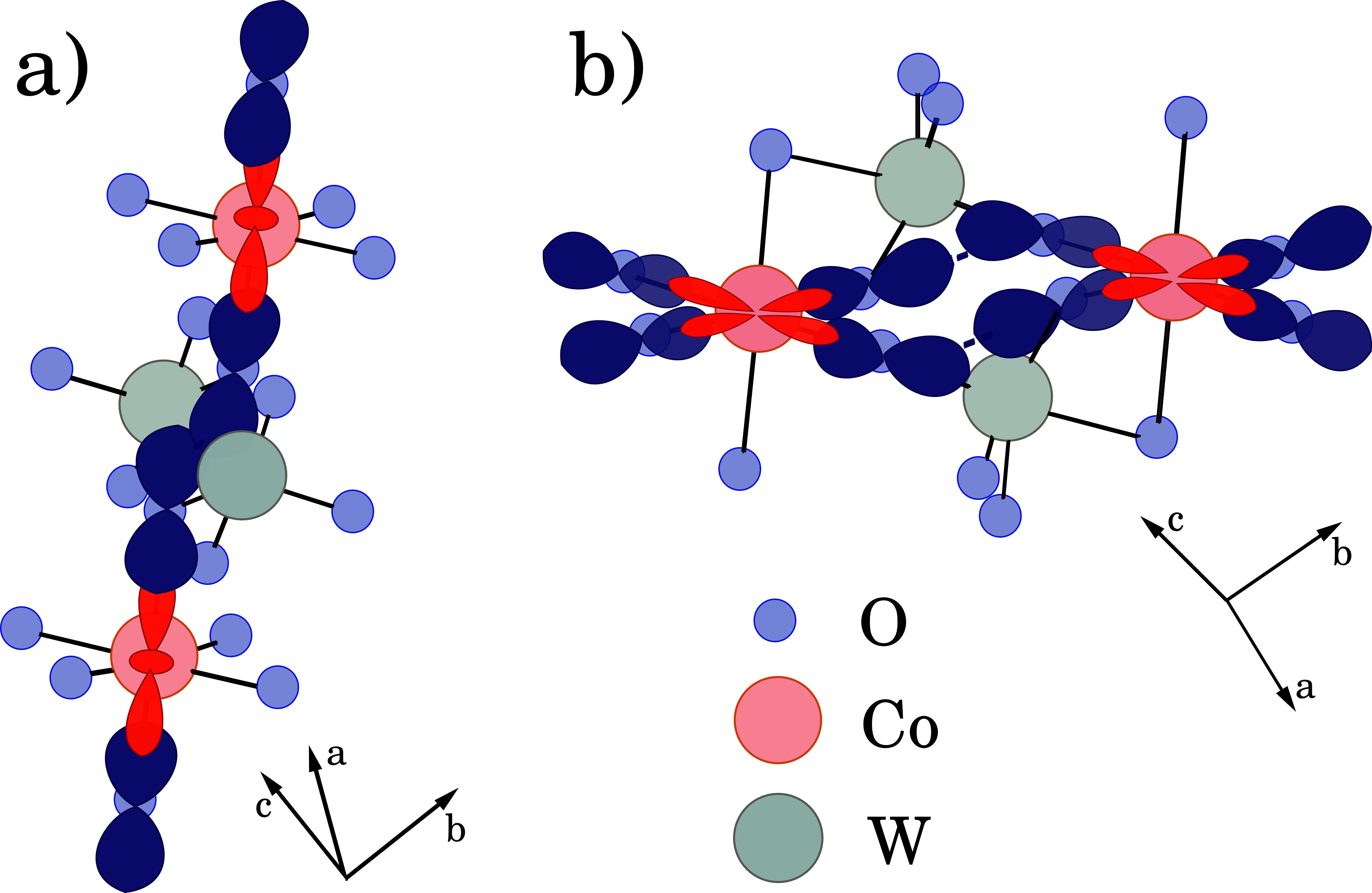}
\end{center}
\caption{\label{pathes}(color online) The schematic representation of the exchange paths for the largest exchange integrals with the Co--Co bond lengths $6.648 \AA$ and $5.650 \AA$.}
\end{figure}

The analysis of the partial contributions shows that the largest
contribution to $J_5$ arises from the overlap between
$3z^2-r^2$ orbitals centered on different sites.
Structurally, it is clearly seen that this  result may be attributed to
the super-superexchange interaction 
via $p$-orbitals of O in the WO$_4$ group, as shown in
Fig.~\ref{pathes}(a). The second strongest exchange integral, $J_3$, belongs to the Co triangular plane that results from the overlap of the $x^2-y^2$ orbitals of Co with the $p$-orbitals of O via the WO$_4$ group, as shown in Fig.~\ref{pathes}(b).

\begin{table*}
\caption{Super-superexchange paths $J_1$, $J_2$, $J_3$, $J_4$ and $J_5$, and their geometrical parameters.}
  \begin{tabular}{c c c c c c c c c}
    \hline \\
    \vspace{1 mm}
     \textbf{Exchange parameters} & \multicolumn{4}{c}{\textbf{Distances \AA}} &\textbf{Co-W-Co angle (deg)} & \multicolumn{2}{c}\textbf{Bond Angles (deg)} & \textbf{Torsion angle (deg)} \\
     \cline{2-5}\cline{7-8}
     & Co-Co & Co-O& O...O & O-Co & & Co-O...O& O...O-Co & Co-O...O-Co\\
\hline
  $J_1$=0.77~$K$, AFM  & 4.907	& 2.187	& 2.663& 2.089  &85.01& 88.53  & 135.11 & 93.72\\
  $J_2$=0.44~$K$, AFM  & 5.618	& 2.189 & 2.929& 2.085  &103.24& 100.52 & 165.39 & 31.75\\
  $J_3$=1.44~$K$, AFM  & 5.650  & 2.084 & 2.737& 2.089  &101.17& 114.048& 155.03 & 41.54\\
  $J_4$=0.55~$K$, AFM  & 5.865	& 2.085	& 2.823& 2.188  &106.40& 144.95 & 126.44 & 11.94\\
  $J_5$=3.0~$K$, AFM   & 6.648 & 2.189 & 2.383& 2.189  &127.70& 158.16 & 158.16 & 180\\

    \hline
 \end{tabular}
\label{Table2}
\end{table*}

According to the Goodenough-Kanamori-Andersen rules (GKA-rules), both of these exchange
constants ($J_3$ and $J_5$) should be AFM,\cite{Goodenough1955,Kanamori1959}  the strong AFM exchange interaction is expected when the partially filled \textit d orbitals overlap with a non-magnetic ion angle of 180$^\circ$, whereas weak ferromagnetic interaction is exhibited when the angle is close to 90$^\circ$.
The other exchange constants are much smaller because the CoO$_6$ octahedra
are isolated from each other by the  Li and W ions. Even for the shortest Co-Co
bond, $4.90$~$\AA$, it is difficult to find a possible exchange paths with the
large overlap of Co-$3d$ orbitals centered on different sites.

 Upon examining the copper oxide compounds with a super-superexchange interaction route, Koo \textit{et al.} proposed that the super-superexchange strength of M-O...O-M  increases with increasing bond angle of M-O...O and decreases with O...O distance.\cite{Koo2005} In particular, the super-superexchange interaction is non-negligible only when the O...O distance is close to or shorter than the van der Waals distance (2.8\AA), and the M-O...O angles are near 160$^\circ$.\cite{Whangbo2003}  The bond lengths and bond angles are summarized in Table~\ref{Table2}.  From Table~\ref{Table2}, we can conclude that $J_5$ should be the largest because of the shortest O...O (2.383\AA) distance and the largest Co-O..O angle of 158.16$^\circ$. Generally, 
the super-superexchange 
via several O ions is often found to be a strong exchange interaction.\cite{Markina2014,Whangbo2003} The second largest exchange interaction is expected to be $J_3$, because of the intermediate O...O distance and Co-O...O and O...O-Co bond angles.  $J_2$ and $J_4$ could be the weakest because the O...O distances are slightly larger than the van der Waals distance of 2.8\AA.  These findings based on bond length and bond angle alone are in good agreement with those obtained from our theoretical calculations shown above.


\subsection{\label{sec:level1} Phase Diagram \protect\\ }

\begin{figure}
\begin{center}
\includegraphics[width=3.5in]{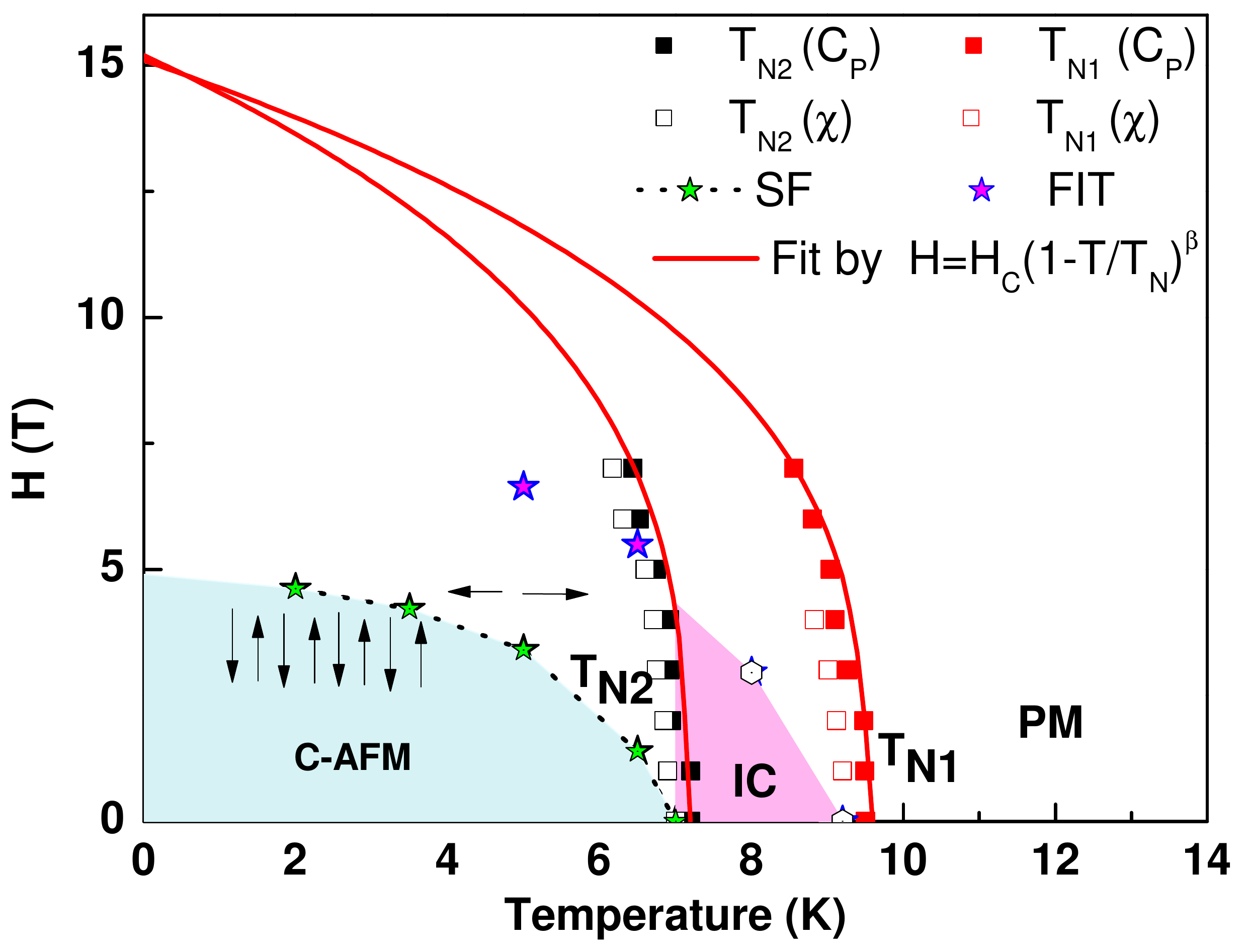}
\end{center}
\caption{\label{fig-phase}(color online)Phase diagram constructed from susceptibility (open symbol), M vs H (star symbols), and specific heat measurements (closed symbol). The Solid lines are power law fitting and the dashed lines are guides for the eye. }
\end{figure}

Since the slope of M(H) isotherms below T$_{N2}$ increases with field, and the C$_P$ peaks at T$_{N2}$ remain sharp to indicate a persistent LRO, the first slope change of M(H) for temperatures below T$_{N2}$ (see Fig.~\ref{fig-MH}) must result from a spin-flop transition.  A spin-flop transition occurs when the applied field is high enough to overcome the on-site spin anisotropy, but the AFM LRO is preserved after all spin are flopped perpendicular to the field direction.  However, as seen in Fig.~\ref{fig-MH}, there exists a second slope change for the M(H) below T$_{N2}$, which cannot be attributed to an additional spin flop transition due to smaller slope, thus a canted ferromagnetic phase transition is possible.

Figure. \ref{fig-phase} shows the H-T phase diagram, constructed from the $\chi$(H,T), M(H,T) and C$_P$(H,T) measurement results, up to 7 T of the VSM field limit. Both spin-flop (SF) and field-induced transitions (FIT) are identified from the M(H) isotherms through their derivatives.  It is not expected for one AFM ordering to have two consecutive spin-flop transitions along one axis through single anisotropy gain. Furthermore, the second slope below T$_{N2}$ is smaller, which goes against the definition of spin-flop transition. However, it is likely that the second slope change for M(H) below T$_{N2}$ indicates an onset of a canted ferromagnetic transition from the AFM phase after all of the spins are flopped.

Using the T$_{N}$(H) values obtained from the C$_P$(T) data at various fields, the critical field H$_c$, and exponent $\beta$ can be calculated through a power low fitting of H/H$_c$=(1-T/T$_c$)$^\beta$. This yields H$_c$ =15.3 T for T$_{N2}$, H$_c$=15.1 T for T$_{N1}$, and $\beta$=1/3 as expected from the mean field theory prediction. Between T$_{N2}$$<$T$<$T$_{N1}$, a single spin-flop transition is identified, which could be attributed to the spin-flop transition of an AFM phase below T$_{N1}$.  Preliminary neutron diffraction studies indicate that the AFM phase below T$_{N1}$ is an incommensurate AFM phase, and the AFM phase below T$_{N2}$ is commensurate, the details of which will be reported elsewhere.\cite{Sunil2014}
\begin{figure}
\begin{center}
\includegraphics[width=3.5in]{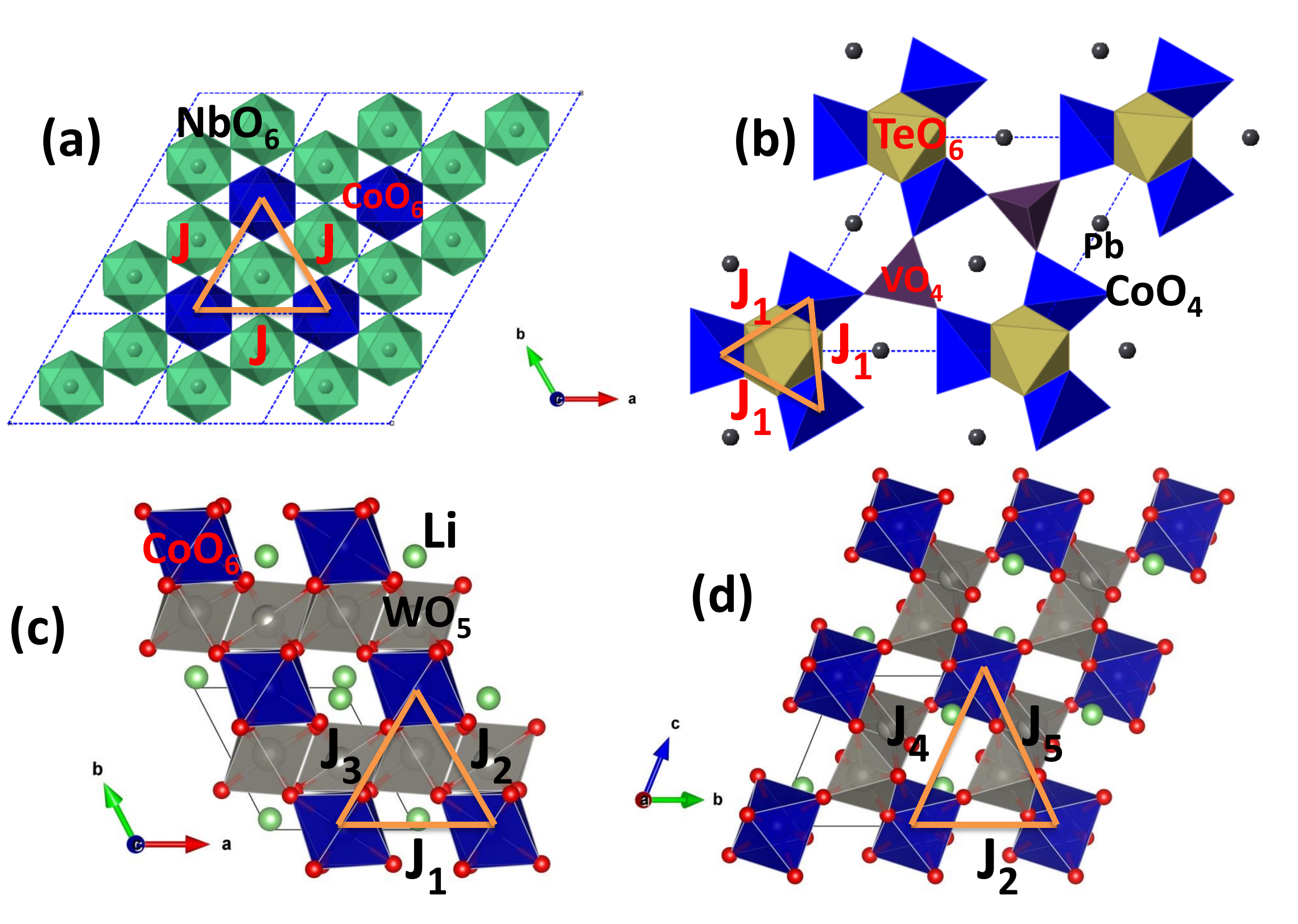}
\end{center}
\caption{\label{fig-threecrystal}(color online) The crystal structure along the \textit {ab} plane for (a) Ba$_3$CoNb$_2$O$_9$ and (b) Pb$_3$TeCo$_3$V$_2$O$_{14}$. (c) $ab$-plane and (d) \textit {bc} planes of Li$_2$Co(WO$_4$)$_2$.  All $J_i$ coupling constants are defined following References \cite{Yokota2014, Markina2014} and Fig.~\ref{fig-structure}.}
\end{figure}


It is worthwhile to compare the H-T phase diagrams of Li$_2$Co(WO$_4$)$_2$ with two other cobaltate systems, Ba$_3$CoNb$_2$O$_9$ and Pb$_3$TeCo$_3$V$_2$O$_{14}$, which exhibit similar two-step AFM transitions.\cite{Yokota2014, Markina2014} The quasi-equilateral triangular arrangements of Co ions in Li$_2$Co(WO$_4$)$_2$, Ba$_3$CoNb$_2$O$_9$, and Pb$_3$TeCo$_3$V$_2$O$_{14}$ are depicted in Fig. \ref{fig-threecrystal}.
The spin structure of Ba$_3$CoNb$_2$O$_9$ can be extracted from the layers of Co spins within the CoO$_6$ octahedra in triangular lattice, and the weak interlayer coupling is obtained through the double corner-sharing NbO$_6$ layers. An easy-axis anisotropy and two-step AFM transition (near T$_{N1}$ = 1.39 and T$_{N2}$ = 1.13 K) in Ba$_3$CoNb$_2$O$_9$ was identified, which strongly suggests that the triangular geometric frustration is lifted via possible magnetophonon coupling, as revealed by the 3D LRO of AFM in steps.
The structure of Pb$_3$TeCo$_3$V$_2$O$_{14}$ consists of quasi-1D structure in which CoO$_6$ octahedra form a unique triangular tubing along the $c$-direction with the two-step AFM transitions found near T$_{N1}$ = 9 K and  T$_{N2}$ = 6 K.  It is noted that the CoO$_6$ trimer within each triangular tubing is coupled through a super-superexchange route via corner-sharing with the TeO$_6$ octahedra along the c-direction, and these CoO$_6$ trimers within the $ab$-plane also form a superlattice of triangular lattices. The stronger intra-chain coupling could lead to the first incommensurate AFM ordering at T$_{N1}$ that eventually orders as a commensurate AFM below T$_{N2}$. Although Li$_2$Co(WO$_4$)$_2$ does not possess a perfect triangular symmetry compared to the other two cobaltates, the triclinic symmetry of the Co spins can be simplified as two quasi-equilateral triangles of $J_1-J_2-J_3$ within the $ab$-plane and $J_2-J_4-J_5$ within the $bc$-plane, as illustrated in Fig.~\ref{fig-structure}, with the Co-Co distances summarized in Table II. The unique coordination between CoO$_6$ bridged with WO$_4$ pairs could also compete with the two nearly orthogonal ($\beta$=91.46$^\circ$) quasi-triangles.  We believe that the common character of the nearly identical two-step AFM transitions found in these three samples may be attributed to the magnetophonon coupling of the bridging polyhedra, which preferably lifts the moderate geometric frustration of the triangular Co spins.

\section{Conclusion}

In summary, Li$_2$Co(WO$_4$)$_2$ exhibits a two-step successive three-dimensional antiferromagnetic transition at T$_{N1}$ $\sim$ 9 K  and T$_{N2}$ $\sim$ 7 K.
The data collected in $\chi$(H,T), M(H,T) and C$_P$(H,T) measurements were used to establish the magnetic phase diagram of Li$_2$Co(WO$_4$)$_2$. This diagram was
compared with that of two other cobaltate systems, Ba$_3$CoNb$_2$O$_9$ and Pb$_3$Co$_3$TeAs$_2$O$_{14}$ with triangular motif of the crystal structure, in which similar two-step AFM transition was found.  The spin frustration ratio, \textit f = $\mid$$\Theta$$\mid$/T$_N$ $\sim$ 3, indicates that the system is moderately frustrated, which could break the triangular symmetry in both $ab$- and $bc$-planes for Co spins in the unique triclinic crystal structure.
The analysis of the results of the LDA+$U$ calculations allowed to find the strongest exchange interactions, which is between fifth nearest neighbors. This is the super-superexchange coupling via two oxygen ions, which results in the formation of AFM chains forming triangular network. Similar long-range exchange interactions were found in other Co$^{2+}$ based systems having two-step AFM transitions: Ba$_3$CoNb$_2$O$_9$ and Pb$_3$Co$_3$TeAs$_2$O$_{14}$.

\section*{Acknowledgments}
The results of experimental measurements were obtained by IPM, RS, GNR, FCC and supported by MOST under project number MOST-102-2119-M-002-004. The results of theoretical calculations: electronic structure and magnetic parameters were obtained by AVU  and SVS and supported by the grant of the Russian Scientific Foundation (project no. 14-22-00004).

\end{document}